\documentclass[aps,prl,reprint,superscriptaddress]{revtex4-1}

\usepackage{graphicx}
\usepackage{amssymb,amsfonts,amsmath,color}
\begin{document}


\title{Dynamic sampling and information encoding in biochemical networks}


\author{Garrett D.\ Potter}
\affiliation{Department of Physics, Oregon State University, Corvallis, OR USA}
\author{Tommy A.\ Byrd}
\affiliation{Department of Physics and Astronomy, Purdue University, West Lafayette IN, USA}
\author{Andrew Mugler}
\affiliation{Department of Physics and Astronomy, Purdue University, West Lafayette IN, USA}
\author{Bo Sun}
\affiliation{Department of Physics, Oregon State University, Corvallis, OR USA}


\begin{abstract}
Cells use biochemical networks to translate environmental information into intracellular responses. These responses can be highly dynamic, but how the information is encoded in these dynamics remains poorly understood. Here we investigate the dynamic encoding of information in the ATP-induced calcium responses of fibroblast cells, using a vectorial, or multi-time-point, measure from information theory. We find that the amount of extracted information depends on physiological constraints such as the sampling rate and memory capacity of the downstream network, and is affected differentially by intrinsic vs.\ extrinsic noise. By comparing to a minimal physical model, we find, surprisingly, that the information is often insensitive to the detailed structure of the underlying dynamics, and instead the decoding mechanism acts as a simple low-pass filter. These results demonstrate the mechanisms and limitations of dynamic information storage in cells.
\end{abstract}


\maketitle

Cells utilize cascades of biochemical pathways in order to translate
environmental cues into intracellular responses
\cite{Sommer2003,Lim2006}. Due to extensive feedbacks and cross-talk
among these signaling pathways
\cite{Subramaniam2005,Pawson2014,Daly2001,Morrison2012}, messenger
molecules exhibit rich dynamic modes, such as waves, oscillations, and
pulses. Recent work in cell biology has suggested a new perspective
in cell signaling: the dynamics, or temporal profiles, of messenger
molecules allow cells to encode and decode even more rich and
complex information than static profiles do \cite{Kholodenko2006,Lahav2013}. For instance,
during inflammation response, exposure to tumor necrosis factor-$\alpha$ (TNF$\alpha$) causes the transcription factor NF-$\kappa$B
to oscillate between the nucleus and cytoplasm of a cell \cite{Baltimore2002}, whereas bacterial lipopolysaccharide (LPS)
triggers a single wave of NF-$\kappa$B within the cell
\cite{Baltimore2005}. Therefore, the dynamics of NF-$\kappa$B encode the
identity of external stimuli. In another example, stimulation of pheochromocytoma
cells (PC12) cells by epidermal
growth factor (EGF) leads to transient mitogen-activated protein kinase (MAPK)
activation and cell proliferation, whereas stimulation by nerve growth factor (NGF) leads to sustained MAPK
activation and cell differentiation \cite{Marshall1995}. These and other examples raise
the question of how one quantifies the information carried by signaling dynamics.

Information theory provides a useful framework to address such
questions \cite{Adami2004,Bialek2012,Nemenman2014}. In the simplest
case, one calculates the scalar mutual information between states
of extracellular stimuli (typically well-controlled
discretized values) and states of the cell (typically protein
concentrations measured at a certain time). Mutual information
characterizes the correlation between environmental cues and cell
responses, and conveniently expresses such correlations in units of
bits. This unified framework has been successfully employed to
quantitatively understand the amount of information that can be
transmitted through a biochemical pathway (channel capacity)
\cite{Levchenko2011}, mechanisms of mitigating errors
\cite{Kuroda2013}, and design principles of signaling network
architectures \cite{Bowsher2014}.

Recently, inspired by the fact that cells utilize dynamic signaling to
encode and decode information, a multivariate, or vectorial, mutual information has
been proposed \cite{Wollman2014}. In this new framework, cellular
responses are described by vectors of dimension $n$, which consist of
cellular states sampled at multiple time points. The vectorial mutual information is generally higher than the scalar
mutual information, indicating that signaling dynamics indeed allow
richer content to be transmitted. It has also been shown that sampling
cellular states at multiple time points eliminates extrinsic noise---noise that degrades information due to cell-to-cell variability.

In light of these results, we ask what is the optimal strategy for
cells to utilize the power of vectorial mutual information? How should a cell sample its own temporal
profiles? Can cells use vectorial mutual information to distinguish
different dynamic states of the underlying signaling pathways? To
address these questions, we combine experimental measurements of
ATP-induced Ca$^{2+}$ responses with theoretical analysis, to
systematically study scalar and vectorial mutual information in a
dynamic signaling system. We find that given different physiological
constraints, the optimal sampling depends on the
starting time, sampling rate and memory capacity. We characterize how vectorial information is affected by intrinsic and extrinsic noise, in both the experimental system and a simple physical model.
Surprisingly, we find that
vectorial mutual information is often insensitive to the detailed structure of the underlying dynamics, failing to distinguish between, for example,
oscillatory and relaxation dynamics. We explain this observation by deriving the connections between vectorial and scalar information, which reveals that in a particular regime vectorial encoding acts as a simple low-pass filter.

\section{Results}
\begin{figure}[t]
\centering 
\includegraphics[width=0.95\columnwidth]{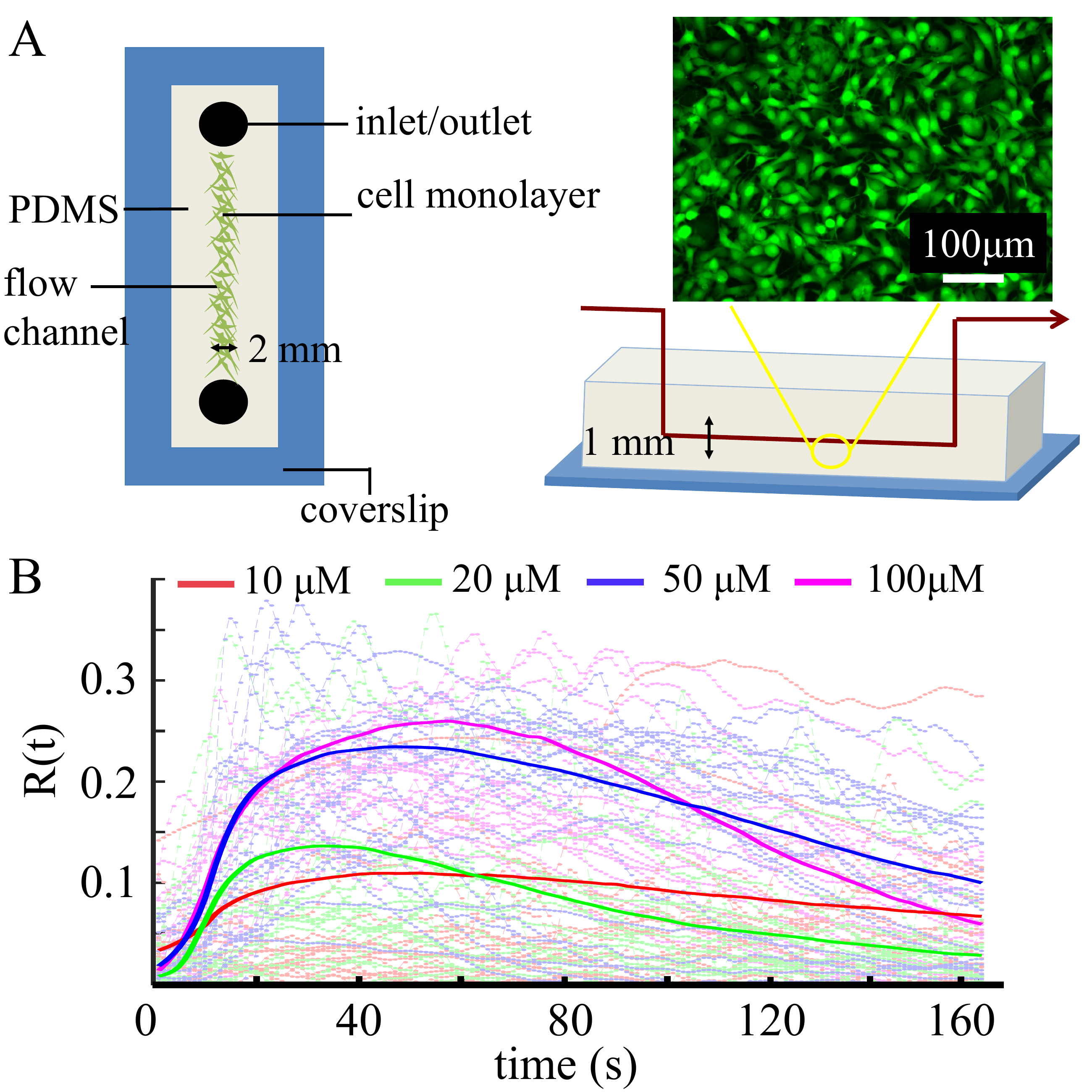}
\caption{Schematics of experimental setup. (A) The top and cross view
  of the microfluidics device to deliver ATP solution to cultured
  fibroblast (NIH 3T3) cells. Inset: fluorescent calcium imaging of a
  typical experiment. (B) Relative fluorescent intensities
  indicating the calcium dynamics $R_i(t)$ of individual cells (dashed
  lines) and their average (solid lines) when stimulated by external
  ATP at four different concentrations.\label{expsetup}}
\end{figure}

To investigate properties of dynamic encoding, we focus on the calcium
dynamics of fibroblast cells in response to extracellular adenosine
triphosphate (ATP), a common signaling molecule involved in a range of
physiological processes such as platelet aggregation \cite{Gachet1999}
and vascular tone \cite{Hochhauser2006}. ATP is detected by P2
receptors on the cell membrane, and triggers the release of second
messenger inositol trisphosphate (IP3). IP3 activates the ion channels
on endoplasmic reticulum (ER) which allows free calcium ions to flux
into the cytosol. The nonlinear interactions of Ca$^{2+}$, IP3, ion
channels and ion pumps generate various types of calcium dynamics
which may lead to distinct cellular functions
\cite{Falcke2004,Lahav2013}.

\subsection{Quantifying information in experimental dynamics}
In order to measure the calcium dynamics of fibroblast cells in
response to external ATP stimuli, we employ microfluidic devices for
cell culture and solution delivery as described previously
\cite{Sun2012,Sun2013b}. In brief, NIH 3T3 cells (ATCC) are cultured
in the PDMS (Polydimethylsiloxane) bounded flow channels as shown in
Fig.\ \ref{expsetup}A. A small fraction of red-fluorescent labeled
MDA-MB-231 cells have also been cocultured with fibroblast cells as a
viability control, and are not included in subsequent analysis. After
attaching the glass bottom for 24 hours, the cells are loaded with
fluorescent calcium indicators (FLUO-4, Thermo Fisher Scientific)
according to the manufacture's protocol. ATP solutions diluted by DMEM
(Dulbecco's Modified Eagle Medium) into 10, 20, 50, and 100 $\mu$M are
sucked into the flow channel with a two-way syringe pump (New Era Pump
Systems Inc.) at a rate of 90 $\mu$L/ min. At the same time, we record
fluorescent images of the cell monolayer at 4 Hz for a total of 10
minutes (Hamamatsu Flash 2.8).

In all the experimental recordings, ATP arrives at approximately
$t=10$ sec, and stays at a constant concentration. Since most responses happen
within 2 minutes, we use the first 160 seconds of recording for
subsequent analysis. The time-lapse images are postprocessed
to obtain the fluorescent intensity $I_i(t)$
of each cell $i$ at a given time $t$. We define the calcium response as
$R_i(t) = [I_i(t)-I_i^r]/I_i^r$, where $I_i^r$ is a reference
obtained by averaging the fluorescent intensity of cell $i$ before ATP
arrives (Fig.\ \ref{expsetup}B).

In order to quantify the information encoded in the calcium dynamics
of fibroblast cells in response to ATP, we have analyzed a total of
more than 10,000 cells over 4 different ATP concentrations (10, 20, 50,
100 $\mu$M) as inputs. With the underlying assumption that each input
appears at probability of $1/4$, the same number of cells are selected
for each ATP concentration. The maximum possible mutual information
between the input and output is therefore $\log_2 4 = 2$ bits.

Denoting the dynamic calcium response as $R^\alpha_i(t)$, where $\alpha = 1, 2, 3,
4$ for each ATP concentration, and $i=1, 2,\dots,N$ for each cell ($N\sim$ 2,500),
the scalar mutual information is defined as
\begin{eqnarray}
MI_s(t) = H\left[\left\{R(t)\right\}\right] - \frac{1}{4}\sum_\alpha
H\left[\left\{R^\alpha(t)\right\}\right],
\end{eqnarray}
where $H$ represents the differential entropy, which we
calculate using the k-nearest neighbor method
\cite{Quesenberry1965,Wollman2014}. The first term is the
unconditioned entropy calculated from cellular responses at time $t$
of all four ATP concentrations. The second term is the average of
differential entropy conditioned at each ATP concentration. The scalar mutual
information $MI_s(t)$
measures how much the entropy in the output (cellular responses) is
reduced by knowledge of the input (ATP concentration).  It is a function of the time $t$ at which we take a snapshot
of the system and evaluate the differential entropy across the ensemble of cells.

The vectorial mutual information is defined as
\begin{equation}
MI_v(t_s) = H\left[\left\{R(\vec{t})\right\}\right] - \frac{1}{4}\sum_\alpha
H\left[\left\{R^\alpha(\vec{t})\right\}\right],
\end{equation}
where $\vec{t} = (t_s, t_s + r^{-1}, t_s + 2r^{-1}, \dots, t_s+T_d)$.
When generalizing to the vectorial
mutual information $MI_v$, one has to
specify not only the sampling start time $t_s$ (equivalent to the time $t$ in the case of $MI_s$), but also the sampling duration $T_d$ and the
sampling rate $r$, which opens the possibility of complex sampling strategies.
In the time between $t_s$ and $t_s+T_d$, a fibroblast cell sampling
its calcium concentration at a rate $r$ accumulates
 a vectorial representation of its calcium dynamics with
vector dimension $n = 1+r T_d$. Since the cell has to store
the vector for further processing, $n$ also represents its memory
capacity.

\subsection{Dynamic encoding increases information}

\begin{figure}[t]
\centering
\includegraphics[width=0.95\columnwidth]{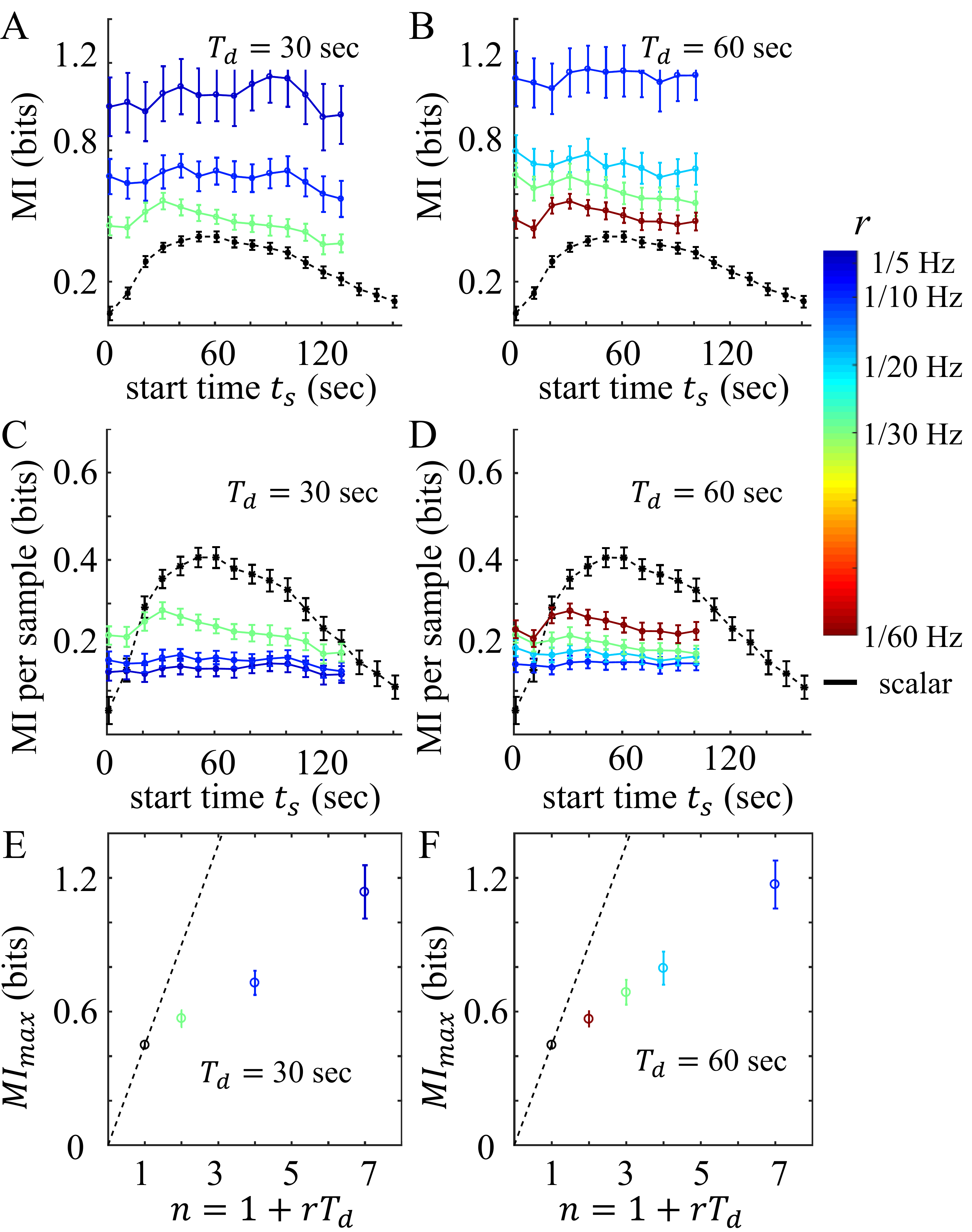}
\caption{Information carried by calcium dynamics of fibroblast
  cells in response to ATP, for fixed sampling duration $T_d$. (A, B)
  Vectorial mutual information $MI_v$ as a function of sampling start time $t_s$ at different sampling rates $r$ (color bar), for (A) $T_d=$ 30 sec and (B) $T_d=$ 60 sec. Black curve is scalar mutual information $MI_s$ at each time point. (C, D) 
  Mutual information per sample for the same conditions as A, B. (E, F) Maximum $MI_v$ over all $t_s$ values, as a function of the memory capacity $n$, for (E) $T_d$ = 30 sec and (F) $T_d$ = 60 sec. Maximum $MI_s$ is plotted at $n=1$. Error bars in A-F represent the  means and standard deviations of 100 bootstraps.
\label{fibroblast_allmicenv_fixduration}}
\end{figure}

We first consider the situation where the sampling duration $T_d$ is fixed.
Fig.\ \ref{fibroblast_allmicenv_fixduration} shows the mutual
information of both scalar ($MI_s$) and vectorial encoding ($MI_v$) from fibroblast
calcium dynamics for $T_d = 30$ sec (A, C, E) and $T_d = 60$ sec (B, D, F). As seen in
Fig.\ \ref{fibroblast_allmicenv_fixduration}A and B, $MI_s$ first rises, then falls, as a function of time. This is due to the separation, then convergence, of the four ATP-conditioned responses as a function of time, as seen in Fig.\ \ref{expsetup}B: better-separated responses contain more information about the ATP level. This shape is also reflected in $MI_v$, with additional smoothing due to the repeated sampling.

Fig.\ \ref{fibroblast_allmicenv_fixduration}A and B also show that $MI_v$ increases with
sampling rate $r$.
This is intuitive, since a larger sampling rate produces a larger number of samples $n = 1 + rT_d$, which increases the amount of information extracted from the dynamics.
While the results in
Fig.\ \ref{fibroblast_allmicenv_fixduration}A and B are
intuitively expected, it is also important to know the efficiency for
dynamic encoding. To this end, we have calculated the mutual
information per sample, defined as $MI_v/n$, as shown in
Fig.\ \ref{fibroblast_allmicenv_fixduration}C and D. It is evident
that higher coding efficiency is achieved at smaller sampling
rate.
This is because when the sampling rate is large, samples are spaced closely in time, and therefore contain increasingly redundant information, which lowers the coding efficiency.
The results shown in Fig.\ \ref{fibroblast_allmicenv_fixduration}C and D suggest that although dynamic encoding mitigates intrinsic
noise, it is not enough to allow $MI_v$ to grow faster than linearly with $n$.
Indeed, scalar encoding generally offers better
efficiency than vectorial encoding: as shown in both
Fig.\ \ref{fibroblast_allmicenv_fixduration}C and D,
$MI_s(t_s)>MI_v(t_s)/n$, except at very early times when the
cellular response has just started.

The results of Fig.\ \ref{fibroblast_allmicenv_fixduration}A-D are summarized in Fig.\ \ref{fibroblast_allmicenv_fixduration}E and F, which plot $MI_{max}$, the maximum mutual information
over all possible sampling start times $t_s$. As seen in
Fig.\ \ref{fibroblast_allmicenv_fixduration}E and F,
$MI_{max}$ monotonically increases with $n$, which shows that dynamic encoding improves the information capture. However,
the increase is sublinear, i.e.\ below the dashed line defined by the scalar mutual information, which shows that the efficiency of dynamic encoding decreases with vector length $n$.
Considering
scalar encoding as the limiting case of $r\rightarrow 0$, we
conclude that as the sampling rate increases, mutual information
increases but the coding efficiency per measurement decreases.

\subsection{Dynamics determine optimal sampling rate}

\begin{figure}[t]
\centering
\includegraphics[width=0.95\columnwidth]{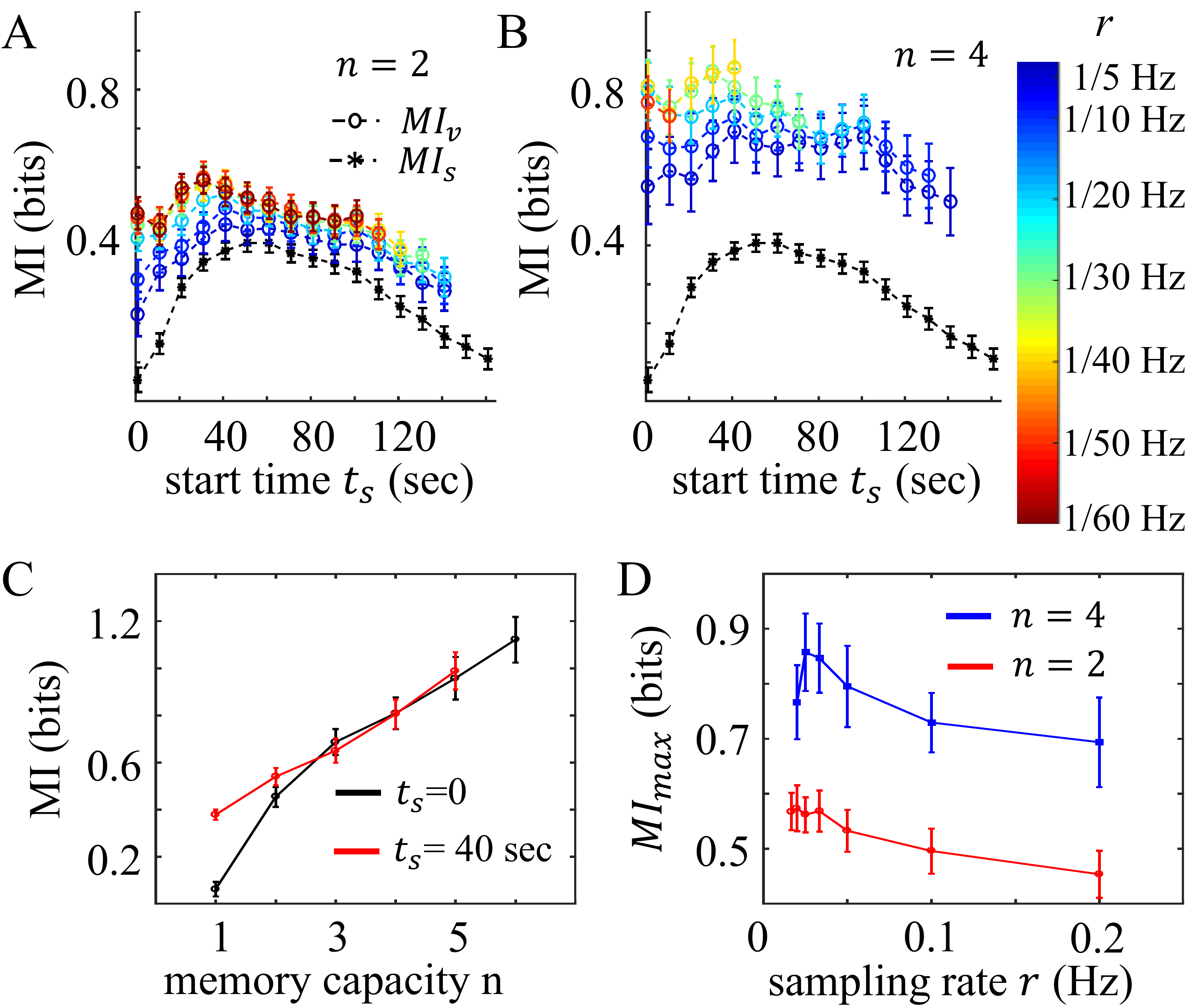}
\caption{Information carried by calcium dynamics of fibroblast
  cells in response to ATP, for a given memory capacity $n$. (A, B) Vectorial mutual
  information $MI_v$ as a function of sampling start
  time $t_s$ at different sampling rates $r$ (color bar) for (A) $n = 2$ and (B) $n = 4$. Black curve
  is scalar mutual information $MI_s$
  each time point. (C) $MI_v$ as a function of
  $n$ at fixed sampling rate $r = 1/30$ Hz and
  sampling start time (black curve $t_s = 0$, red curve $t_s = 40$
  sec). $n=$1 corresponds to $MI_s$. (D) Maximum $MI_v$ over all $t_s$ values, as a function of $r$, for
  fixed memory capacity (red $n=2$, blue $n=4$). Error bars in A-D represent the means and standard
  deviations of 100 bootstraps.
\label{fibroblast_allmicenv_fixcapacity}}
\end{figure}

Cells have limited ability to process dynamically encoded
information. It is conceivable that a biochemical signaling network
processing a vectorial code of high dimension will be complex and
expensive, because it requires a high memory capacity $n$ for storage
and transfer. Therefore a relevant question is, what sampling strategy
can a cell apply when the memory capacity is fixed? Fig.\ \ref{fibroblast_allmicenv_fixcapacity}A and B show
the mutual information as a function of sampling start time $t_s$ when the memory capacity $n$
is fixed, while the sampling rate $r$, and therefore the duration $T_d = (n-1)/r$, are allowed to vary.
Comparing Fig.\ \ref{fibroblast_allmicenv_fixcapacity}A to B, we see that larger memory capacity $n$ generally allows more information to be
transmitted, as was the case in Fig.\ \ref{fibroblast_allmicenv_fixduration}. This trend is quantified in Fig.\ \ref{fibroblast_allmicenv_fixcapacity}C, which plots the mutual information as a function of $n$, for fixed sampling rate and at two particular starting times $t_s$. We see that the amount of
information significantly depends on $t_s$ for small $n$, while the
difference diminishes at larger $n$. This is because information is upper-bounded (at 2 bits in
our case), which requires that all curves, regardless of $t_s$, ultimately saturate with increasing $n$.
Therefore we see that larger
memory capacity not only encodes higher information, but also helps cells to obtain more uniform readouts.

We also see in Fig.\ \ref{fibroblast_allmicenv_fixcapacity}A and B that for a given
$n$, there is an optimal sampling rate $r$ that maximizes the information. This is made more evident by considering, as before, the maximum mutual information $MI_{max}$ over all possible start time $t_s$, which is plotted as a function of $r$ in 
Fig.\ \ref{fibroblast_allmicenv_fixcapacity}D. Particularly, for $n=4$ (blue curve), we see that $MI_{max}$ is maximal at a particular sampling rate. This is because, for a fixed number of samples $n$, sampling too frequently results in redundant information, as discussed above; while sampling too infrequently places samples at late times, when the dynamic responses have already relaxed (see Fig.\ \ref{expsetup}B).
Therefore it is generally
beneficial to sample at a lower rate except when the sampled points
are too far apart, which places samples outside the `high
yield' temporal region.
The tradeoff between these two effects leads to the optimal sampling rate, where the information gathered is the largest.

\subsection{Vectorial information is insensitive to detailed dynamic structure}

Is vectorial encoding sensitive to the underlying details of the dynamic response?
In order to answer this question, and to provide a mechanistic understanding of dynamic information
transmission in biochemical networks, we construct a minimal stochastic model with the aim of recapitulating the key
features of the fibroblast response. As a minimal model we consider a
damped harmonic oscillator in a thermal bath, driven out of equilibrium by a
time-dependent forcing $F(t)$. The magnitude of the external forcing is
proportional to a scalar input, which is analogous to the ATP
concentration. The displacement of the particle $x(t)$, like the
calcium dynamics, can then be analyzed to infer the information that the oscillator encodes about the
input.

\begin{figure}
\centering
\includegraphics[width=0.95\columnwidth]{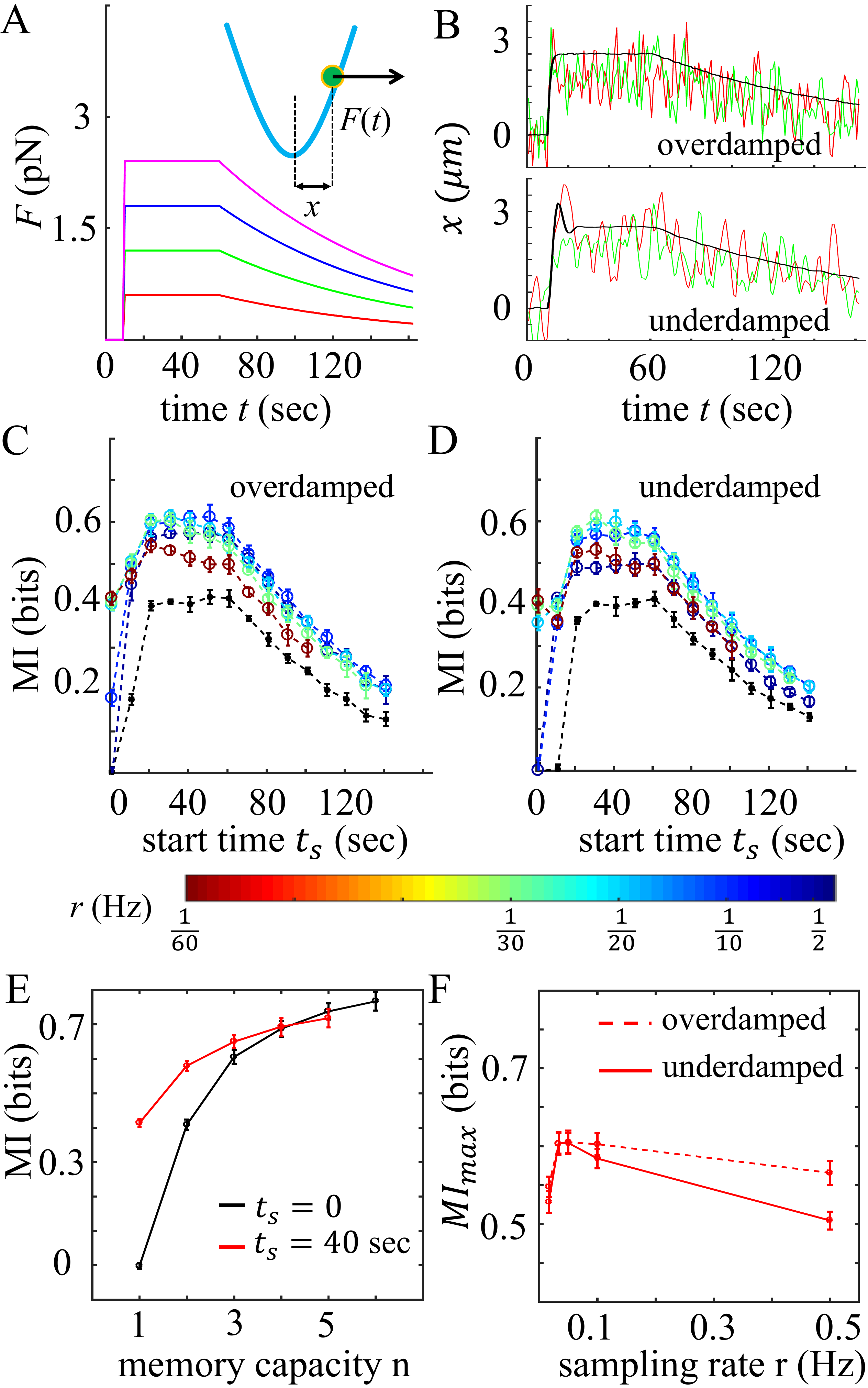}
\caption{Information encoding in the noisy harmonic oscillator model.
(A) Oscillator at position $x(t)$ is subjected to random thermal forcing as
  well as deterministic forcing $F(t)$ (Eq.\ \ref{oscillator}). Four force magnitudes $g_{1,2,3,4} = \left\{\text{0.6 pN, 1.2 pN, 1.8 pN, 2.4 pN}\right\}$ serve as input, while $x(t)$ is output. Other parameters are $t_1 = 10$ sec, $t_2 = 60$ sec, and $\beta = 0.01$ sec$^{-1}$.  (B) Two sample trajectories (red and green curves) and the average of 5,000
trajectories (black curves) corresponding to $g_4 = 2.4$ pN. Upper: overdamped oscillators with $m = 0.4 m_c$, where $m_c = \gamma^2/(4k) = 0.25$ mg is the mass at critical damping, $\gamma  = 1$ pN$\cdot$sec/$\mu$m, and $k=$ 1 pN/$\mu$m. Lower: underdamped oscillators with $m=9m_c$. Other parameters are $k_BT$ = 0.5 pN$\cdot\mu$m and $\delta k = 0.2$ pN/$\mu$m.
(C, D) Vectorial mutual
  information $MI_v$ as a function of sampling start
  time $t_s$ at different sampling rates $r$ (color bar) and memory capacity $n=2$, for (C) overdamped and (D) underdamped oscillators. Black curve
  is scalar mutual information $MI_s$
  each time point. (E) $MI_v$ as a function of
  $n$ at fixed sampling rate $r = 1/30$ Hz and
  sampling start time (black curve $t_s = 0$, red curve $t_s = 40$
  sec). $n=$1 corresponds to $MI_s$.  
 (F) Maximum $MI_v$ over all $t_s$ values, as a function of $r$, for
  fixed memory capacity $n=2$. Error bars in C-F represent the means and
  standard deviations of 20 independent trials each.
\label{noisyoscillator}}
\end{figure}

The equation of motion for the oscillator is given by
the Langevin equation \cite{Kampen2007}
\begin{eqnarray}
\label{oscillator}
m\frac{d^2x}{dt^2} + \gamma\frac{dx}{dt} + kx &=& g_\alpha F(t) + \psi(t), \nonumber \\
\langle\psi(t)\psi(t')\rangle &=& 2k_BT\gamma\delta(t-t'), \nonumber \\
F(t) &=&
\begin{cases}
0 & t < t_1 \\
1 & t_1 \le t \le t_2 \\
e^{-\beta(t-t_2)} & t > t_2.
\end{cases}
\end{eqnarray}
Here $m$ is the mass, $\gamma$ is the drag coefficient, and $k$ is the spring constant.
$\psi(t)$ is the random forcing arising from thermal fluctuations with energy $k_BT$; it is Gaussian and white, and represents intrinsic noise. The form of the external forcing $F(t)$, illustrated in Fig.\ \ref{noisyoscillator}A for four magnitudes $g_{1,2,3,4}$, is chosen to reflect the fact that following initial elevation, cells relax to their resting level of cytosolic calcium concentration at the end of experimental recording (see Fig.\ \ref{expsetup}B).
To account for the extrinsic noise observed in our cellular
system \cite{potter2015collective}, we have allowed the spring constant for each oscillator trajectory to
vary uniformly around $\langle k\rangle$ with a standard deviation $\delta k$.
Fig.\ \ref{noisyoscillator}B shows sample trajectories $x(t)$ for two cases: when the oscillations are overdamped ($m < m_c$) or underdamped ($m > m_c$), where $m_c \equiv \gamma^2/(4k)$ is the mass at critical damping.

Fig.\ \ref{noisyoscillator}C and D show, for the overdamped and underdamped cases,
the scalar and vectorial mutual information between input $g_\alpha$ and output $x(t)$, as a
function of sampling start time $t_s$, for various sampling rates $r$ and fixed memory capacity $n=2$.
Additionally, Fig.\ \ref{noisyoscillator}E shows the mutual information as a function of $n$ at fixed $r$ for the overdamped case, while Fig.\ \ref{noisyoscillator}F shows the maximum mutual information as a function of $r$ at fixed $n$ for both cases. Comparing Fig.\ \ref{noisyoscillator} to Fig.\ \ref{fibroblast_allmicenv_fixcapacity}, we see that our minimal model is sufficient to capture the key features of the experiments. Specifically, comparing Fig.\ \ref{noisyoscillator}C and D to Fig.\ \ref{fibroblast_allmicenv_fixcapacity}A, we see that the model captures the non-monotonic shape of the mutual information as a function of start time $t_s$, as well as the improvement of vectorial encoding (colors) over scalar encoding (black). Comparing Fig.\ \ref{noisyoscillator}E to Fig.\ \ref{fibroblast_allmicenv_fixcapacity}C, we see that the model captures the increase of mutual information with memory capacity $n$, as well as the large-$n$ convergence of curves with different $t_s$. Finally, comparing Fig.\ \ref{noisyoscillator}F to Fig.\ \ref{fibroblast_allmicenv_fixcapacity}D, we see that the model captures the presence of an optimal sampling rate $r$ that negotiates the tradeoff between samples that are well-separated, yet confined to the high-yield region ($t_1\leq t \leq t_2$ in the model). These correspondences validate the model, and allow us to use the model to ask how vectorial encoding depends on the structure of the underlying dynamic responses.



The noisy oscillator model allows us to explore two qualitatively different regimes of dynamic structure. In the overdamped regime, the  thermal noise overpowers the oscillations, and the dynamics are dominated by fluctuations (Fig.\ \ref{noisyoscillator}B, upper). In contrast, in the underdamped regime, the oscillations overpower the thermal noise, and the dynamics are dominated by the underlying oscillatory structure (Fig.\ \ref{noisyoscillator}B, lower). Since vectorial mutual information corresponds to sampling the dynamics at regular intervals, it is natural to hypothesize that the amount of information extracted from underdamped dynamics will be higher than that extracted from overdamped dynamics, because underdamped dynamics have a more ordered structure. Fig.\ \ref{noisyoscillator}C and D compare the mutual information in the overdamped and underdamped cases. Surprisingly, we see that the amounts of information are roughly equivalent in the two cases. It is evident from Fig.\ \ref{noisyoscillator}C and D that the equivalence holds at varying sampling rates $r$ and start times $t_s$ (including the start time at which the information is maximal, Fig.\ \ref{noisyoscillator}F). In particular, the equivalence holds when the sampling rate $r$ equals the oscillation frequency of the underdamped oscillator, $\nu = \sqrt{(k/m)(1-m_c/m)}/(2\pi) \approx 1/10$ Hz in Fig.\ \ref{noisyoscillator}. This is true despite the fact that $r=\nu$ is where we would have expected the vectorial information to benefit most from sampling the periodic dynamics instead of noisy dynamics. We have also checked that the equivalence holds for a large range of intrinsic and extrinsic noise levels. The previously demonstrated correspondence between the model and the experiments suggests that in the fibroblasts as well, ordered dynamics would not provide more information than noisy dynamics, at least as quantified by the vectorial mutual information. We expand upon this conclusion in the Discussion.

%

\subsection{Differential effects of intrinsic and extrinsic noise}

\begin{figure}[t]
\centering
\includegraphics[width=0.75\columnwidth]{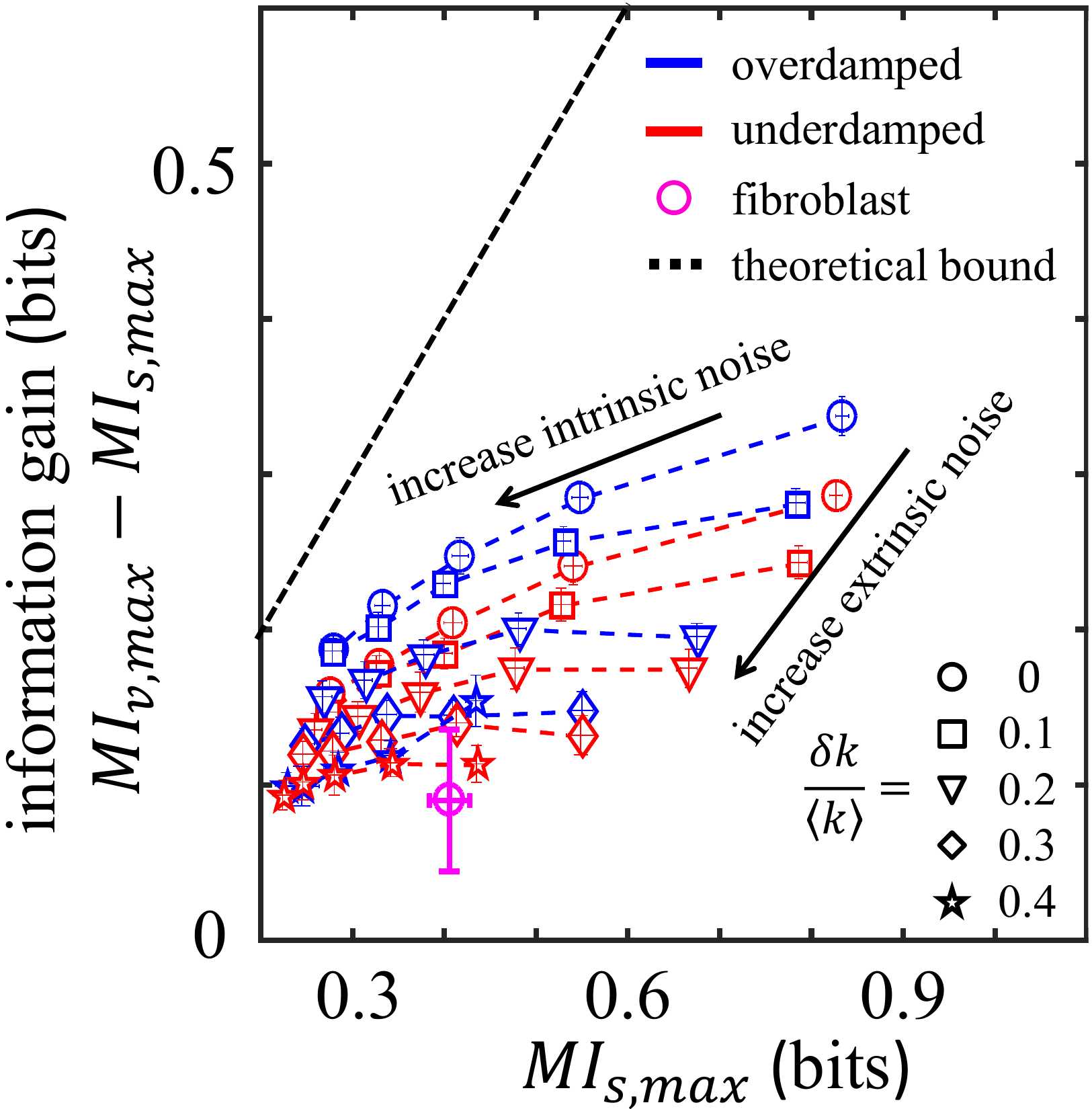}
\caption{Gain of vectorial over scalar mutual information, where each is maximized over start times $t_s$. For vectorial information, memory capacity is $n=2$ and sampling rate is $r=1/10$ Hz. Fibroblast data are compared against over- and underdamped oscillator models and detailed model of calcium dynamics \cite{potter2015collective}. In oscillator model, parameters are as in Fig.\ \ref{noisyoscillator}, and intrinsic noise is governed by $k_BT$, which varies from $0.2$ to $1$ pN$\cdot\mu$m, while extrinsic noise is governed by $\delta k/\langle k\rangle$, which varies from $0$ to $0.4$. In calcium model, extrinsic noise is governed by parameter fold-change factor $F$ \cite{potter2015collective}. Error bars
  represent means and standard deviations of 100 bootstraps (fibroblast data), 20 independent trials of 5,000
  trajectories each (oscillator model). 
\label{informationgain}}
\end{figure}

Vectorial mutual information $MI_v$ is larger than scalar mutual information
$MI_s$ in part because repeated sampling helps to mitigate intrinsic
noise \cite{Wollman2014}. Yet, in the case of the fibroblast cells, the gain of $MI_v$ over
$MI_s$ is often small. For example, as seen in Fig.\ \ref{fibroblast_allmicenv_fixcapacity}A, at $n=2$ and $r=
1/10$ Hz, whereas $MI_s$ can be as large as $\sim$$0.4$ bits, the further increase of $MI_v$ over $MI_s$ is less than $\sim$$0.1$ bits. We make this observation quantitative by defining the information gain $MI_{v,max}-MI_{s,max}$, where each is maximized over the start time $t_s$. Fig.\ \ref{informationgain} shows the information gain vs.\ $MI_{s,max}$ for the fibroblasts at $n=2$ and $r=1/10$ Hz (pink circle). The fact that the gain is small ($0.1$ bits) suggests that additional factors, apart from intrinsic noise alone, reduce the efficacy of vectorial encoding.

To explore this hypothesis in a systematic way, we again turn to our minimal oscillator model. For both the overdamped and underdamped oscillator, we compute $MI_{s,max}$ and the information gain. In the model, the intrinsic noise is governed by the thermal energy $k_BT$. The model also provides an opportunity to investigate the effects of extrinsic noise, which is governed by $\delta k/\langle k\rangle$, the relative width of the distribution of spring constants.
As shown in Fig.\ \ref{informationgain}, when the intrinsic noise increases while the extrinsic noise is fixed, both the scalar information and the information gain decrease, as expected (dashed lines). The decrease in scalar information is more pronounced than the decrease in the gain, which is consistent with the fact that vector information is beneficial for mitigating intrinsic noise. On the other hand, when the extrinsic noise increases while the intrinsic noise is fixed, the gain decreases more rapidly, while the scalar information decreases less rapidly (the different symbols). This implies that the gain is more sensitive to extrinsic noise than intrinsic noise.

In the context of the fibroblast population, these results suggest
that extrinsic noise (cell-to-cell variability), not intrinsic noise, is primarily responsible for degrading the performance of vectorial encoding and producing small information gains.

\subsection{Redundant information and low-pass filtering}

The vectorial mutual information $MI_v$ can never be larger than the sum of the scalar mutual information values $MI_s(t_i)$ taken individually at each time point $t_i$. The reason is that there will always be some nonnegative amount of redundant information between the output at one time and the output at another time. Denoting the redundant information as $MI_{red}$, we formalize this statement as
\begin{equation}
\label{bound}
MI_{red} = \left[\sum_{i=1}^n MI_s(t_i)\right] - MI_v = n \langle MI_s \rangle - MI_v \ge 0,
\end{equation}
where as before $t_i = t_s + i r^{-1}$, and in the second step we rewrite the sum in terms of the temporal average $\langle MI_s \rangle = n^{-1} \sum_{i=1}^n MI_s(t_i)$. In the limit that the dynamics are approximately stationary, such as in the high-yield regions of Figs.\ \ref{expsetup}B and \ref{noisyoscillator}B, $MI_s$ is approximately independent of time, and $\langle MI_s \rangle = MI_s$. For $n=2$, as in Fig.\ \ref{informationgain}, Eq.\ \ref{bound} then becomes
\begin{equation}
\label{bound2}
MI_v-MI_s \le MI_s.
\end{equation}
Eq.\ \ref{bound2} expresses the intuitive fact that the gain upon making an additional measurement can never be more than the information from the original measurement, for a stationary process. Eq.\ \ref{bound2} is plotted in Fig.\ \ref{informationgain} (dash-dotted line), and we see that it indeed bounds all data from above, as predicted.

\begin{figure}[t]
\centering
\includegraphics[width=0.75\columnwidth]{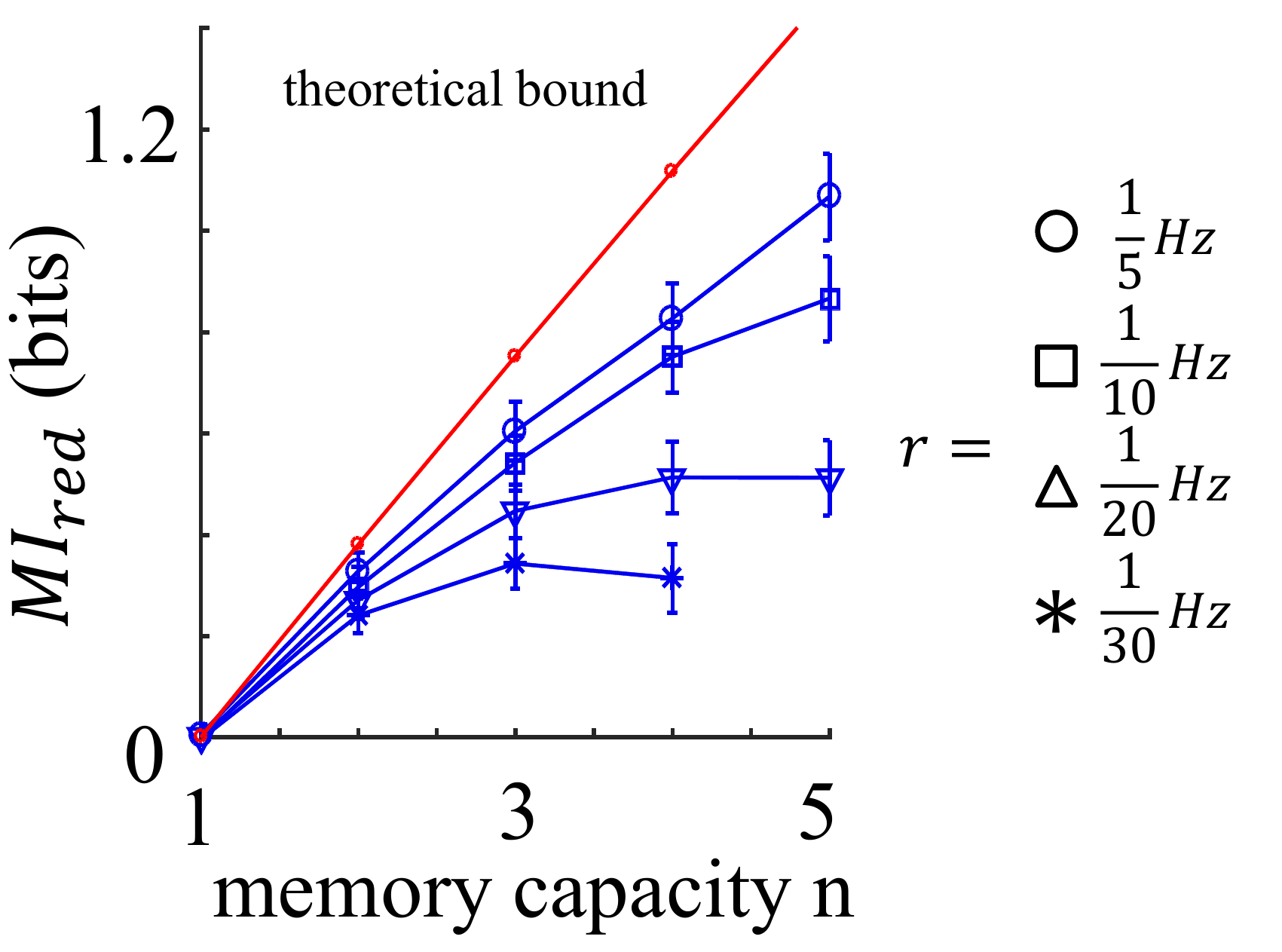}
\caption{Redundant information of dynamic encoding for fibroblast
  cells. Redundant information $MI_{red}$ (Eq.\ \ref{bound}) is plotted as a function of memory capacity $n$ for
  varying sampling rates $r$ and compared with the theoretical bound (Eq.\ \ref{bound3}).
\label{reducedinformation}}
\end{figure}

The redundant information in Eq.\ \ref{bound} can be directly measured in the experiments. Fig.\ \ref{reducedinformation}A shows the redundant information in the fibroblast calcium dynamics as a function of the memory capacity $n$, computed from the scalar and vectorial mutual information according to Eq.\ \ref{bound}.
We see that the redundant information depends on the sampling rate $r$ (symbols) and appears to be bounded from above by a roughly linear function of $n$. Can we understand this dependence theoretically? To address this question, we return to Eq.\ \ref{bound}. We rearrange Eq.\ \ref{bound} as $MI_{red} = (n-1) \langle MI_s \rangle - \Delta$, where we define $\Delta = MI_v - \langle MI_s \rangle$. Since the vectorial information is not smaller than the scalar information corresponding to any of its time points, it is also not smaller than the average scalar information. Therefore, $\Delta \ge 0$, and we have
\begin{equation}
\label{bound3}
MI_{red} \le (n-1) \langle MI_s \rangle
\end{equation}
Eq.\ \ref{bound3} is a linear function of $n$, weakly modified by the fact that $\langle MI_s \rangle$ itself depends on $n$ since it is computed for varying numbers of time points.
Eq.\ \ref{bound3} is compared with the data in
Fig.\ \ref{reducedinformation}, and we see that it indeed predicts the bound well.
Eq.\ \ref{bound3} makes another prediction, namely that the
bound is reached for a stationary process when $\Delta = MI_v - \langle MI_s \rangle \to 0$, i.e.\ when the vector information provides vanishing improvement over the average scalar information. We expect this situation to occur in the limit of large sampling frequency $r$, when samples occur in close succession and offer little additional information beyond a single, scalar measurement. Indeed, we see from the data in Fig.\ \ref{reducedinformation} that consistent with this prediction, the bound is approached in the limit of increasing $r$.

Clearly, the benefit of vectorial encoding is largest when the redundant information is small (the lowest data points in Fig.\ \ref{reducedinformation}). In the ideal case, there is no redundant information at all, and Eq.\ \ref{bound} becomes
\begin{equation}
\label{lowpass}
MI_v = \sum_{i=1}^n MI_s(t_i).
\end{equation}
Here, we see that the vectorial information is simply the sum of the scalar information at each time point. In this sense, Eq.\ \ref{lowpass} describes a low-pass filter: vectorial encoding captures the temporal accumulation of scalar information, as long as the sampling is sufficiently slow to remove the redundancy. Therefore, in this limit the vectorial information records only the slow (low-frequency) variations in the dynamics. This feature may help explain the previous counterintuitive result that the vectorial information is insensitive to the detailed dynamic structure, as we expand upon in the Discussion.

\section{Discussion}
The dynamic waveforms of signaling molecules have offered a new
perspective to understand cellular information encoding. Indeed,
dynamic encoding, as quantified by the vectorial mutual information
$MI_v$, has larger channel capacity than the static encoding, as
quantified by the scalar mutual information $MI_s$ \cite{Wollman2014}. From both
experimental data and a minimal model we presented here, we find that 
dynamic encoding has several key advantages over static
encoding. First, the maximal vectorial information is larger than the maximal scalar information, suggesting that dynamic
encoding provides a more reliable readout of environmental inputs than static encoding does. Second,
while the scalar information can vary significantly with sampling time, the vectorial information is more uniform across sampling start times, even with small vector
dimensions (Fig. \ref{fibroblast_allmicenv_fixcapacity}C and
Fig. \ref{noisyoscillator}E). 

However, the benefit of dynamic encoding comes with the cost of
increasing the memory capacity $n$ of cells. For a fixed memory
capacity, we have shown that the best strategy for cells to adopt is to
sample as slowly as possible while keeping their samples within a ``high-yield'' region, where the mean dynamics depend significantly on the input. Nonetheless, we find that within this region, the benefit of dynamic encoding can depend very little on the detailed structure of the dynamics (persistent oscillation vs.\
monotonic relaxation). Moreover, the gain of dynamic encoding over static encoding can be small, largely due to the presence of extrinsic, as opposed to intrinsic, noise.

The finding that vectorial information is largely insensitive to the detailed dynamics is surprising, and is likely a reflection of the type of dynamics we investigate here, as well as the vectorial measure itself. To accurately model the experimental dynamics, we have considered noisy dynamics arising from a driven oscillator. Although this has allowed us to probe both noise-dominated and oscillation-dominated regimes, these dynamics remain mean-reverting and confined to a stationary or cyclo-stationary state. It is likely that other classes of dynamics, such as temporal ramps, would emerge as having uniquely higher vectorial information than stationary dynamics. Furthermore, the vectorial information itself, as defined here, reports correlations between a categorial input variable and a regularly sampled output trajectory. It is likely that more sophisticated information-theoretic measures would be more sensitive to dynamic details, such as the mutual information between input and output trajectories, which has been argued to play a biological role in cell motility \cite{tostevin2009mutual, tostevin2010mutual}.

Our results suggest that dynamic and
static encoding mechanisms are deeply connected. By
introducing the redundant information $MI_{red}$, we have made this connection rigorous.
Specifically, combining Eqs.\ \ref{bound} and \ref{bound3} yields $\langle MI_s
\rangle \le MI_v \le n \langle MI_s\rangle$, which shows explicitly that
the vectorial information is bounded from both above and
below by quantities determined by the window-averaged scalar information $\langle MI_s\rangle$.
Taking a window average of the scalar information is equivalent to the downstream network acting as a low-pass filter, accumulating temporal measurements at sufficiently low frequencies.
We find that such low-pass
filtering effects are evident from both the experimental and modeling
results.

In this study we have taken the approach that an understanding of both the static and dynamic encoding behaviors
of the fibroblast cells can be obtained from a model based on noisy
harmonic oscillators. Despite the simplicity of the model, we find that
it reproduces the experimental results very well. The
agreement between the experiment and this simple model highlights our central conclusion: the vectorial
mutual information is intrinsically connected with the scalar mutual
information and therefore has limited capability to distinguish underlying
dynamics. Because the model is minimal, we anticipate that it can be extended to answer more general questions about information encoding on a large, multicellular scale. This
is particularly desirable as understanding collective information
processing is a new frontier in systems biophysics \cite{Sun2012,
  Sun2013b, potter2015collective, mugler2016limits, ellison2016cell, camley2015emergent}.

\bibliography{vectorencoding}

\end{document}